\documentclass[a4paper,11pt]{article}
\usepackage{epsfig}
\usepackage{cite}

\begin{document}

\newcommand{\lst}       {\mbox{$\ell^*$}}
\newcommand{\est}       {\mbox{$\rm e^*$}}
\newcommand{\mst}       {\mbox{$\mu^*$}}
\newcommand{\tst}       {\mbox{$\tau^*$}}
\newcommand{\llst}      {\mbox{$\ell^*\ell^*$}}
\newcommand{\eest}      {\mbox{$\rm e^*e^*$}}
\newcommand{\mmst}      {\mbox{$\mu^*\mu^*$}}
\newcommand{\ttst}      {\mbox{$\tau^*\tau^*$}}
\newcommand{\sqrts}     {\mbox{$\sqrt{s}$}}
\newcommand{\roots}     {\sqrts}
\newcommand{\cm}        {centre-of-mass}
\newcommand{\rvis}      {\mbox{$R_{\rm vis}$}}
\newcommand{\epair}     {\mbox{$\rm e^+e^-$}}
\newcommand{\mpair}     {\mbox{$\mu^+\mu^-$}}
\newcommand{\tpair}     {\mbox{$\tau^+\tau^-$}}
\newcommand{\qpair}     {\mbox{$\rm q\bar{q} $}}
\newcommand{\w}         {\mbox{$\rm W^\pm$}}
\newcommand{\z}         {\mbox{$\rm Z^0$}}
\newcommand{\g}         {\mbox{$\gamma$}}
\newcommand{\eg}        {\mbox{$\rm e\gamma$}}
\newcommand{\eeg}       {\mbox{$\rm ee\gamma$}}
\newcommand{\mmg}       {\mbox{$\mu\mu\gamma$}}
\newcommand{\ttg}       {\mbox{$\tau\tau\gamma$}}
\newcommand{\llg}       {\mbox{$\rm \ell\ell\gamma$}}
\newcommand{\eegg}      {\mbox{$\rm ee\gamma\gamma$}}
\newcommand{\mmgg}      {\mbox{$\mu\mu\gamma\gamma$}}
\newcommand{\ttgg}      {\mbox{$\tau\tau\gamma\gamma$}}
\newcommand{\llgg}       {\mbox{$\rm \ell\ell\gamma\gamma$}}

\newcommand{\etal}       {{\it et al.}}
\newcommand{\PLB}[3]     {Phys.\ Lett.\ {\bf B#1} (#2) #3}
\newcommand{\ZPC}[3]     {Z.\ Phys.\ {\bf C#1} (#2) #3}
\newcommand{\NIMA}[3]    {Nucl.\ Instr.\ Meth.\ {\bf A#1} (#2) #3}
\newcommand{\PPE}[1]     {CERN--PPE/97--{#1}}
\newcommand{\EP}[1]      {CERN--EP/98--{#1}}
\newcommand{\PRLD}[3]    {Phys.\ Rev.\ Lett.\ {\bf D#1} (#2) #3}
\newcommand{\PRL}[3]     {Phys.\ Rev.\ Lett.\ \textbf{#1} (#2) #3}
\newcommand{\PRD}[3]     {Phys.\ Rev.\ {\bf D#1} (#2) #3}
\newcommand{\PTP}[3]     {Prog.\ Theor.\ Phys.\ \textbf{#1} (#2) #3}
\newcommand{\SJNP}[3]    {Sov.\ J.\ Nucl.\ Phys.\ \textbf{#1} (#2) #3}
\newcommand{\NPB}[3]     {Nucl.\ Phys.\ {\bf B#1} (#2) #3}
\newcommand{\CPC}[3]     {Comp.\ Phys.\ Comm.\ {\bf #1} (#2) #3}
\newcommand{\EPJC}[3]     {Eur.\ Phys.\ J.\ {\bf C#1} (#2) #3}

\newcommand{\meegg}  {103.2}    
\newcommand{\mmmgg}  {103.2}
\newcommand{\mttgg}  {103.2}
\newcommand{\lastcm} {208.3}

\topsep0pt plus 1pt

\begin{titlepage}
\begin{center}{\large   EUROPEAN ORGANIZATION FOR NUCLEAR RESEARCH
}\end{center}\bigskip
\begin{flushright}
       CERN-EP/2002-043   \\ 18 June 2002
\end{flushright}
\bigskip\bigskip\bigskip\bigskip\bigskip
\begin{center}{\huge\bf   Search for Charged Excited Leptons \\ in
{\boldmath $\rm e^+e^-$} Collisions at {\boldmath $\sqrt{s}$}~=~183-209~GeV
}\end{center}\bigskip\bigskip\bigskip
\begin{center}{\LARGE The OPAL Collaboration
}\end{center}\bigskip\bigskip\bigskip
\bigskip\begin{center}{\large  Abstract}\end{center}
A search for charged excited leptons decaying into a lepton and a
photon has been performed
using approximately 680 $\rm pb^{-1}$ of \epair\ collision data
collected by the OPAL detector 
at LEP at \cm\ energies between 
183~GeV and 209~GeV.  No evidence for their existence was found.  
Upper limits on the product of the cross-section and the branching fraction
are inferred.
Using results from the search for singly produced excited leptons, 
upper limits on the ratio of the excited lepton coupling constant to
the compositeness scale  
are calculated.  
From pair production searches, 95\%~confidence level lower limits on
the masses of excited electrons, muons and taus
are determined to be \meegg~GeV.

\bigskip\bigskip\bigskip\bigskip
\bigskip\bigskip
\begin{center}{\large
(To be submitted to Physics Letters B)
}\end{center}
\end{titlepage}
\begin{center}{\Large        The OPAL Collaboration
}\end{center}\bigskip
\begin{center}{
G.\thinspace Abbiendi$^{  2}$,
C.\thinspace Ainsley$^{  5}$,
P.F.\thinspace {\AA}kesson$^{  3}$,
G.\thinspace Alexander$^{ 22}$,
J.\thinspace Allison$^{ 16}$,
P.\thinspace Amaral$^{  9}$, 
G.\thinspace Anagnostou$^{  1}$,
K.J.\thinspace Anderson$^{  9}$,
S.\thinspace Arcelli$^{  2}$,
S.\thinspace Asai$^{ 23}$
% ...
}\end{center}
\bigskip\bigskip
$^{  1}$School of Physics and Astronomy, University of Birmingham,
Birmingham B15 2TT, UK
\newline
$^{  2}$Dipartimento di Fisica dell' Universit\`a di Bologna and INFN,
I-40126 Bologna, Italy
\newline
$^{  3}$Physikalisches Institut, Universit\"at Bonn,
D-53115 Bonn, Germany
\newline
% ...
$^{  5}$Cavendish Laboratory, Cambridge CB3 0HE, UK
\newline
% ...
$^{  9}$Enrico Fermi Institute and Department of Physics,
University of Chicago, Chicago IL 60637, USA
\newline
% ...
$^{ 16}$Department of Physics, Schuster Laboratory, The University,
Manchester M13 9PL, UK
\newline
% ...
$^{ 22}$Department of Physics and Astronomy, Tel Aviv University,
Tel Aviv 69978, Israel
\newline
$^{ 23}$International Centre for Elementary Particle Physics and
Department of Physics, University of Tokyo, Tokyo 113-0033, and
Kobe University, Kobe 657-8501, Japan
\newline
% ...
$^{  e}$ and Department of Experimental Physics, Lajos Kossuth University,
 Debrecen, Hungary
\newline
% ...

\section{Introduction}
\label{section:intro}
Models in which fermions have
substructure attempt to
explain, among other things, the well-ordered pattern of fermion
generations observed in nature.
The existence of excited states of the Standard Model
fermions would be a natural consequence of fermion compositeness.
Excited leptons 
could be produced in \epair\ collisions and are
expected to decay via the emission of a
gauge boson (\g, \z\ or \w)~\cite{bib:old_theory_paper}.   

This paper presents results from a search for excited electrons (\est), 
muons (\mst) and tau leptons (\tst) decaying electromagnetically,
using data collected by the OPAL experiment at LEP.  
Results presented in this paper are obtained using a larger data
sample, as well as a significantly improved analysis, compared to our
previous publications~\cite{bib:ex_opal_results}.
Searches for excited leptons have also been performed by other LEP
collaborations~\cite{bib:lep} and by the HERA
experiments in electron-proton collisions~\cite{bib:hera}.

At LEP, excited leptons could be produced in pairs or in
association with a 
Standard Model lepton.  Both single and pair production of excited
leptons proceed through {\it s}-channel photon and \z\ diagrams.  
In addition, {\it t}-channel photon and \z\ exchange diagrams contribute to the
single and pair production of excited electrons.  
The {\it t}-channel contribution, although expected to be negligible for
pair production, causes excited electrons to be
singly produced predominantly in the forward region with the recoiling
electron outside the detector acceptance.
Thus, in addition to final states containing two leptons and one or
two photons (\llg, \llgg), a separate search for the single production
of excited electrons 
with an undetected electron (\eg) is also performed.

The results presented in this paper are interpreted in the context of
the framework described in \cite{bib:ex_theory1,bib:ex_theory2}. 
In this phenomenological model, 
the interaction between excited leptons and a gauge boson 
($\ell^*\ell^*V$), which largely determines the cross-section for pair
production of excited leptons, is vector-like.
The single production cross-section and branching
fractions of excited leptons are determined by the strength of 
the $\ell\ell^*\rm V$ coupling.  This interaction can be     
described by the following SU(2)$\times$U(1) gauge invariant effective 
Lagrangian~\cite{bib:ex_theory1,bib:ex_theory2}
\[
  {\cal L}_{\ell\ell^* V} =
  \frac{1}{2 \Lambda} \bar{\ell}^*\sigma^{\mu\nu}
  \left[g f \frac{ \mbox{\boldmath $\tau$} }{2}
    \mbox{\boldmath {\rm\bf W}}_{\mu\nu} +
  g^\prime f^\prime \frac{Y}{2} B_{\mu\nu} \right] \ell_{\rm L} +
  {\rm hermitian~conjugate},
\]
\noindent
where $\sigma^{\mu\nu}$ is the covariant bilinear tensor, 
\mbox{\boldmath $\tau$}
denotes the Pauli matrices, $Y$ is the weak hypercharge, ${\bf W_{\mu\nu}}$ 
and $B_{\mu\nu}$ represent the Standard Model gauge field tensors and the couplings $g,g^\prime$ are the SU(2) and U(1) 
coupling constants of the Standard Model.
The compositeness scale is set by the parameter $\Lambda$ which has units 
of energy.  Finally, the strength of the $\ell\ell^*\rm V$ coupling is 
governed by the constants $f$ and $f^\prime$.
These constants can be interpreted as weight factors associated with 
the different gauge groups. 
The values of $f$ and $f^\prime$ dictate the relative branching
fractions of excited leptons to each gauge boson. 
The branching fraction of electromagnetically decaying excited charged
leptons is significant for most values of $f$ and
$f^\prime$ except in the case where $f=-f^\prime$ which entirely
forbids this particular decay. 
As a result of the clean characteristic signatures expected,
the photon decay constitutes one of the most sensitive channels for the
search for excited leptons, even for values of $f$ and $f^\prime$ for which other
decay modes dominate.
To reduce the number of free parameters it is customary to assume
either a relation between $f$ and $f^\prime$ or set one coupling to zero.
For easy comparison with previously published results, limits calculated
in this paper correspond to the coupling choice $f=f^\prime$.  This assignment
is a natural choice which forbids excited neutrinos from decaying
electromagnetically.  
For this particular coupling choice, the electromagnetic branching
fraction of charged excited leptons drops smoothly from 100\% for
masses below the \z\ and \w\ mass thresholds to about 30\% for masses in
excess of 200~GeV.

\section{Data and Simulated Event Samples}

The data analysed were collected by the OPAL
detector~\cite{bib:opal}  
at \cm\ energies ranging from 183~GeV to 209~GeV during the LEP runs
in the years 1997 to 2000.   
The search for excited leptons is based on a total of 684.4~$\rm
pb^{-1}$ of data for which all relevant detector components were
fully operational.  
For the purpose of accurately interpreting the results in terms of
limits on excited lepton masses and couplings,
the data are divided into 16 \cm\ energy bins
analysed separately.  The energy range, luminosity
weighted mean \cm\ energy and integrated luminosity of each bin are
summarised in Table~\ref{table:bin}. 
The uncertainty on the measured beam energy is
approximately 25~MeV~\cite{bib:LEP_energy} and is correlated between
\cm\ energy bins.  
In addition to the high energy data, approximately 10~$\rm pb^{-1}$ of
calibration data collected in 1997-2000 at a \cm\ energy near the
\z\ mass were used to study the detector response.

\begin{table}
\begin{center}
\begin{tabular}{|c|c|r|}\hline
\sqrts\ bin range & $<\sqrt{s}>$  &
\multicolumn{1}{|c|}{$\cal L$} \\  
\multicolumn{1}{|c}{(GeV)} & \multicolumn{1}{|c}{(GeV)} & \multicolumn{1}{|c|}{$\rm (pb^{-1})$} \\ \hline\hline
178.00 - 186.00 & 182.7 & 63.8 \\ \hline
186.00 - 190.40 & 188.6 & 183.2 \\ \hline
190.40 - 194.00 & 191.6 & 29.3 \\ \hline
194.00 - 198.00 & 195.5 & 76.5 \\ \hline
198.00 - 201.00 & 199.5 & 76.9 \\ \hline
201.00 - 203.75 & 201.9 & 44.5 \\ \hline
203.75 - 204.25 & 203.9 & 1.5 \\ \hline
204.25 - 204.75 & 204.6 & 9.7 \\ \hline
204.75 - 205.25 & 205.1 & 60.0 \\ \hline
205.25 - 205.75 & 205.4 & 3.6 \\ \hline
205.75 - 206.25 & 206.1 & 14.3 \\ \hline
206.25 - 206.75 & 206.5 & 107.3 \\ \hline
206.75 - 207.25 & 206.9 & 5.7 \\ \hline
207.25 - 207.75 & 207.5 & 0.5 \\ \hline
207.75 - 208.25 & 208.0 & 7.2 \\ \hline
$>$ 208.25        & 208.3 & 0.5 \\ \hline\hline
\multicolumn{2}{c|}{} & 684.4 \\ \cline{3-3}
\end{tabular}
\end{center}
\caption{\label{table:bin}
The luminosity weighted mean \cm\ energy and integrated luminosity of
each energy bin.} 
\end{table}

Distributions of kinematic variables and selection efficiencies for
excited leptons were modelled using samples of simulated events
obtained using the EXOTIC~\cite{bib:exotic} Monte Carlo event generator.
The matrix elements~\cite{bib:ex_theory1,bib:exotic_matrix}
implemented in EXOTIC include all the spin correlations in the
production and decay of excited leptons.

The Standard Model processes at different \cm\ energies 
were simulated using a variety of Monte Carlo event generators. 
Bhabha events were simulated using the BHWIDE~\cite{bib:bhwide} and
TEEGG~\cite{bib:teegg} generators, muon and tau pair events
using both KORALZ~\cite{bib:koralz} and KK2F~\cite{bib:kk2f}, 
$\rm e^+e^-\rightarrow\rm q\bar{\rm q}(\gamma)$ events using
PYTHIA~\cite{bib:pythia} and KK2F, four fermion processes using 
KORALW~\cite{bib:koralw} and grc4f~\cite{bib:grc4f}, di-photon
production using RADCOR~\cite{bib:radcor}, and
two-photon events
($\rm
e^+e^-$~$\rightarrow$~$e^+e^-\gamma\gamma$~$\rightarrow$~$e^+e^-f\bar{f}$)   
using  VERMASEREN~\cite{bib:vermaseren},
PHOJET~\cite{bib:phojet} and HERWIG~\cite{bib:herwig}.  Each of the simulated
event samples was processed through the OPAL detector simulation
program~\cite{bib:gopal} and analysed in the same way as data.

\section{Event Selection and Kinematic Fits}
\label{section:selection}

Events are reconstructed from tracks and energy clusters 
defined by requirements similar to those described in
\cite{bib:tracks}.
The background from multihadronic events is substantially reduced by
requiring at least one but no more than six tracks in an event.
Furthermore, the ratio of the number of good tracks, as defined in
\cite{bib:tracks}, to
the total number of tracks reconstructed in the detector
must be greater than 0.2 in order to reduce background from beam-gas 
and beam-wall collisions.  Cosmic ray events are suppressed using
information from the time-of-flight counters and the central tracking
chamber~\cite{bib:cosmic}. 

Tracks and energy clusters in an event are 
grouped into jets using a cone algorithm~\cite{bib:cone} 
with a cone half-angle of 0.25~radians and minimum jet energy of 2.5~GeV.
The parameters defining a jet were chosen
to maximise the signal efficiency over the broadest possible range of 
excited lepton masses and \cm\ energies. 
Events are required to contain between two and four jets.  
Jets are classified as leptons or photons using the 
criteria described below applied in the same order as given in the text. 

Photon candidates must have a minimum energy deposited in
the electromagnetic calorimeter equivalent to 5\% of the
beam energy.  A photon jet must either contain no tracks or be
identified as a photon conversion using a neural network
technique~\cite{bib:idncon_nn8}.  Jets in which the
most energetic track has a neural network output greater than 0.9 and
the energy deposited in the hadronic calorimeter is less than 10\% of the
beam energy  
are defined to be photon conversions.  All photon candidates
must lie within $|\cos\theta| < 0.9$
 to avoid poorly modelled regions of the detector~\footnote{The OPAL
coordinate system is defined to be 
right-handed, with the z-axis pointing along the electron beam
direction and the x-axis pointing toward the centre of the LEP ring.
Thus the polar angle $\theta$ used in this paper refers to the angle
with respect to the electron beam direction and the azimuthal angle
$\phi$, the angle measured with respect to the x-axis.}.
The energy and
direction of each photon candidate are determined from the energy and
position of the energy cluster in the electromagnetic calorimeter.

Muon candidates are jets containing exactly one track with associated
hits in the muon detectors or hits in the hadronic calorimeter
consistent with the particle being a muon~\cite{bib:ww}.  
Muons, unlike electrons and photons, do not shower while traversing
additional material present in the forward region of the detector.
Muon candidates are thus allowed to lie within a larger angular acceptance
of $|\cos\theta|<0.95$.  The direction of each muon candidate is given by
the polar and azimuthal angles of the track and the momentum is
calculated from the track curvature and polar angle.

A jet is identified as an electron if it contains exactly one track satisfying
one of the following two requirements:  
the ratio of the electromagnetic energy to the track momentum ({\it
E/p}) lies between 0.8 and 1.4 or the track has an output
greater than 0.9 from a neural network developed to identify
electrons~\cite{bib:idncon_nn8}.
Electron candidates are also required to lie
within $|\cos\theta|<0.9$ to avoid poorly modelled regions of the detector.
The energy of each electron candidate is taken to be the energy deposited in
the electromagnetic calorimeter while the direction is given by 
the polar and azimuthal angles of the track.

Finally, unidentified jets containing at least one track
and lying within $|\cos\theta|<0.95$ are considered to be tau candidates.
Jets in the region $0.90<|\cos\theta|<0.95$ which would satisfy the electron or
photon requirements are discarded from the sample of tau candidates. 
Jets identified as tau candidates are mostly
hadronically decaying taus. 
Tau leptons decaying leptonically are
tagged as electrons or muons by the criteria described
above.  The polar and azimuthal angles of tau candidates are given by
the axis of the jet, corrected for double-counting of tracks and
energy clusters~\cite{bib:MT}.

The different selections used to identify the final states of interest
are described in the following sections and the results are summarised
in Table~\ref{table:results}. 

\subsection{Selection of \boldmath \llgg\ Final States}
\label{section:llgg}
Events containing two lepton candidates of the same flavour and two
identified photons are considered as candidate events for the pair
production of excited leptons.  In addition, events containing two
leptons of different flavours and two photons are considered as
excited tau candidates.  
In order to reduce the background from Standard Model
processes, additional selection criteria are applied to the different
types of candidate events.

\begin{figure}
\begin{center}
\mbox{\epsfig{file=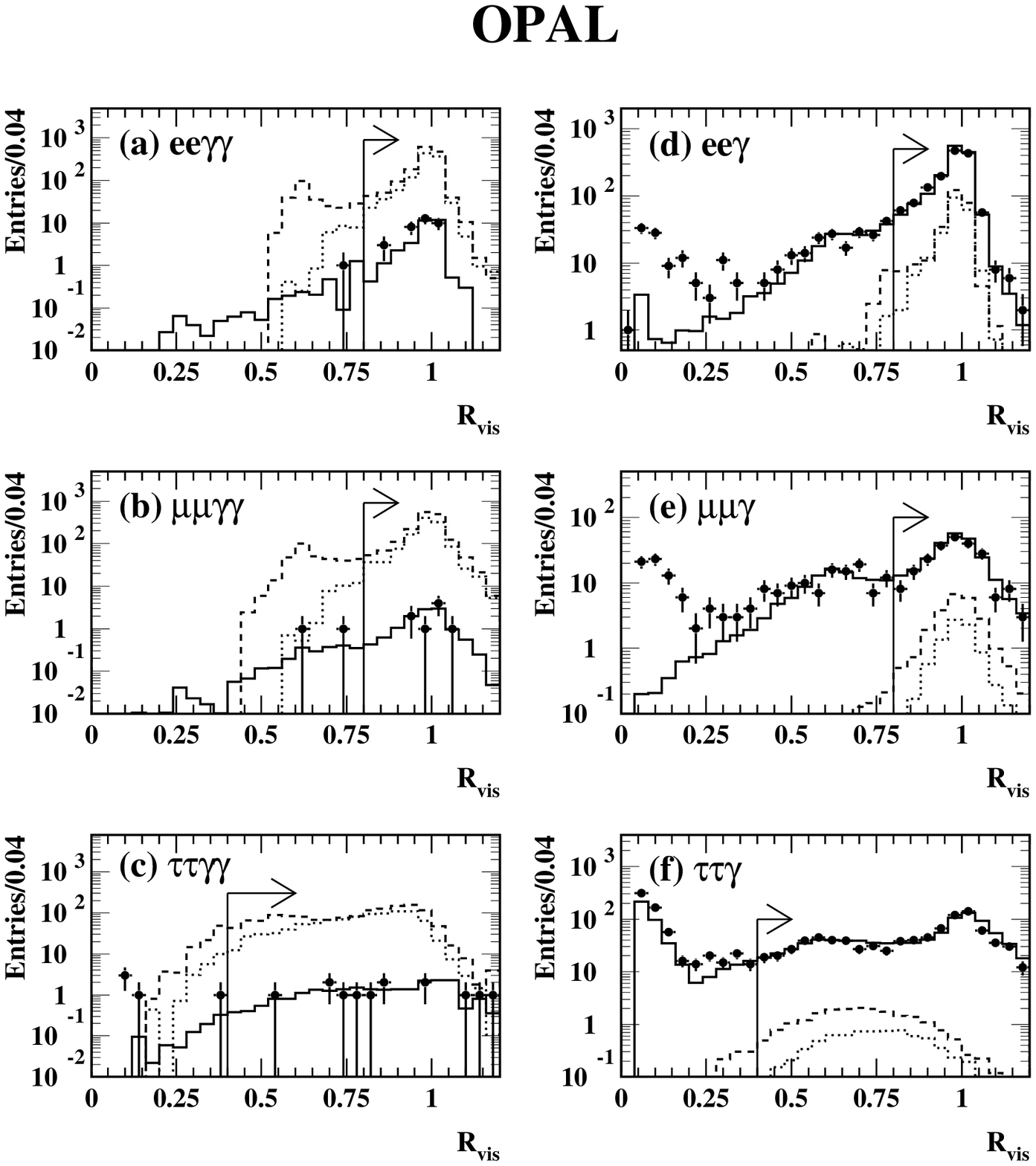,height=17cm}}
\caption{\label{fig:rvis} 
Distributions of the sum of the energies of the two leptons and one or
two photons divided by the \cm\ energy for pair (a-c) and single
(d-f) production candidate events after the preselection.  
The points represent the combined data from all \cm\ energies considered  
while the solid lines are the total expected background from Standard
Model processes.  
The dashed lines represent an example of \lst\lst\ (a-c) and \lst (d-f)
signal events with arbitrarily chosen masses of 40~GeV and 90~GeV
respectively.   The dotted lines show the expected distributions for
pair and singly produced excited leptons with masses of 90~GeV and
180~GeV, respectively.
Distributions of excited lepton signal events are normalised to a
ratio of the coupling constant to the compositeness scale of $\rm 1\
TeV^{-1}$.  
The arrows indicate the accepted regions. 
The background modelling for low values of \rvis\ is discussed in
Section~\ref{section:llgg}.} 
\end{center}
\end{figure}

The quantity \rvis\ is defined to be the sum of the energy of
the particles considered for a given event final state, divided by the
\cm\ energy.  This quantity is required to 
exceed 0.8 for \eegg\ and \mmgg\ candidates and 0.4
for \ttgg\ candidates.  
This criterion reduces the background from two-photon events.  It also
decreases the contamination from \qpair\ events in the \ttgg\ sample.
Figure~\ref{fig:rvis}(a-c) shows the \rvis\ distributions obtained using
the entire data set for each type of candidate event.
The observed discrepancy at small values of \rvis\ in the \ttgg\
sample corresponds to a region where the background is dominated by
two-photon events and does 
not affect the analysis as the events of interest lie in a region of
\rvis\ that is well modelled.  A similar mis-modelling is present
in distributions of $\ell\ell\gamma$ and $\rm e\gamma$ candidate events.  

The remaining background in the
\eegg\ and \mmgg\ samples comes almost entirely from 
\epair\ and \mpair\ events with additional photons.
The background in the \ttgg\ sample consists mostly of 
\tpair\ events with more than one radiated photon,
and a small fraction of \qpair\ events.

\subsection{\boldmath Selection of \llg\ Final States}
\label{section:llg}
Events containing two lepton candidates of the same flavour and at
least one identified photon are considered as candidate events for the
single production of excited leptons.   
In addition, events with two leptons of different flavours and at
least one photon are considered as excited tau candidates.
If more than 
one photon is identified in the event, the most energetic
photon is chosen and the other photon is ignored.  
Events selected as candidates for the pair production of excited
leptons are also considered as single production
candidates.  

To reduce the background from two-photon events the quantity \rvis\ must be 
greater than 0.8 for \eeg\ and \mmg\
candidates, and greater than 0.4 for \ttg\ candidates.  The \rvis\
distributions of each type of candidate event are shown in
Figure~\ref{fig:rvis}(d-f) for data from all the \cm\ energies combined.

\begin{figure}
\begin{center}
\mbox{\epsfig{file=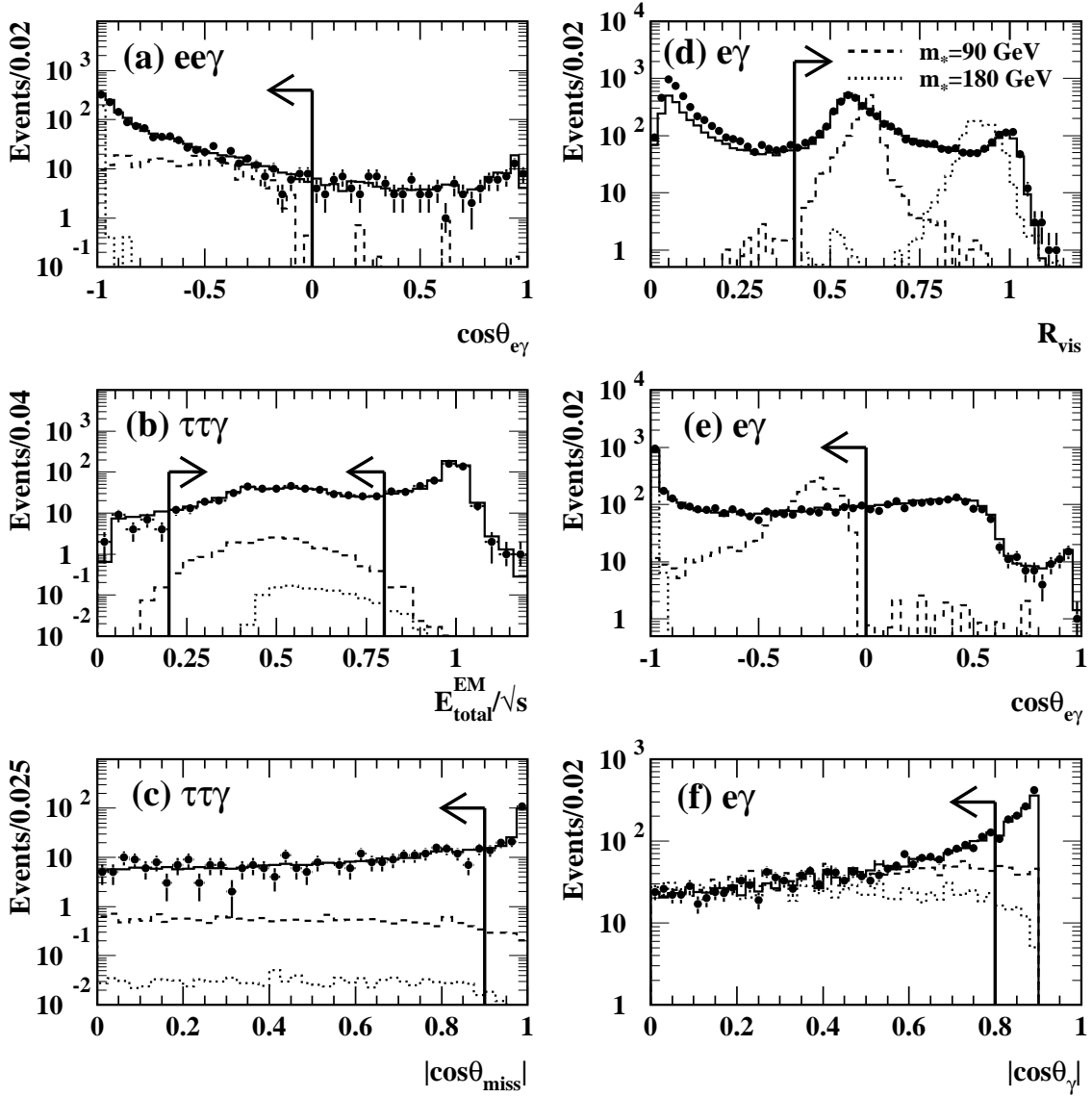,height=17cm}}
\caption{\label{fig:other} Distributions of selection variables for
different final states: (a) cosine of the angle between the most
energetic electron and the photon in \eeg\ events, (b) total energy
deposited in the electromagnetic calorimeter scaled by the \cm\ energy
for \ttg\ events, (c) absolute value of the cosine of the missing
momentum vector polar angle for
\ttg\ events, (d) sum of the energy of the electron and photon divided
by the \cm\ energy for \eg\ events, (e) cosine of the angle between
the electron and photon for \eg\ events and (f) absolute value
of the cosine of the photon polar
angle in \eg\ events.
All selection cuts have been applied, in the same order as described in
the text, up to that on the variable plotted.
The dashed and dotted lines represent examples of
excited lepton signal events with arbitrarily chosen masses of 90~GeV
and 180~GeV, respectively, and branching fraction calculated assuming
a ratio of the coupling constant to the
compositeness scale of $\rm 1\ TeV^{-1}$.   }
\end{center}
\end{figure}

The dominant \epair\ background in the \eeg\ final state
is reduced by requiring that 
the angle between the most energetic electron and photon
($\theta_{e\gamma}$) be greater than $90^0$.  The $\cos\theta_{\rm
e\gamma}$ distribution obtained using data
from all \cm\ energies combined is shown in Figure~\ref{fig:other}(a).

Background from both \epair\ and \mpair\ events in the \ttg\ sample
is reduced by requiring the total energy deposited
in the electromagnetic calorimeter to
be between 20\% and 80\% of the \cm\ energy. 
Finally, the polar angle of the missing momentum vector for the particles
considered in the \ttg\ final state must lie
within $|\cos\theta_{\rm miss}| < 0.9$.  This requirement reduces the
contamination from \qpair\ events.  Figures~\ref{fig:other}(b,c)
show both the total electromagnetic energy and $|\cos\theta_{\rm
miss}|$ distributions of \ttg\ events before applying each cut.
 
After this selection, the remaining background in the \eeg\ and \mmg\ 
samples consists almost entirely
of \epair\ and \mpair\ events with an additional photon.
The background in the \ttg\ sample consists mostly of \tpair\ events
with one radiated photon 
and a small fraction of radiative \epair, \mpair\ and \qpair\ events.

\subsection{\boldmath Selection of $\rm e\gamma$ Final State}
A separate selection for events with one electron and one photon
was developed to increase the efficiency
of the search for singly produced excited electrons where one electron 
travels in the forward region outside the detector acceptance.
 
Candidate events are required to contain at least one photon and at least 
one electron candidate.  Additional jets, if present, are ignored. 
Since the \eeg\ and \eg\ final states are combined to 
calculate a limit on the single production of excited electrons, it is
important to ensure that events are not double-counted.
All events that are selected by the set of general
requirements discussed at the beginning of Section~\ref{section:selection},
but that fail the \eeg\
selection are considered as possible \eg\ candidates.

To reduce the two-photon background, the quantity \rvis\ must satisfy
\rvis\ $> 0.4$.  The angle between the 
electron and photon 
($\theta_{e\gamma}$) is also required to be greater than $90^0$.      
Further reduction of the dominant \epair\ background is
achieved by requiring that the measured polar angle of the photon
satisfies $|\cos\theta_{\gamma}| < 0.8$ and by rejecting events where
the photon is identified as a conversion.  

Figures~\ref{fig:other}(d-f) show the
\rvis, $\cos\theta_{e\gamma}$ and $|\cos\theta_{\gamma}|$ 
distributions of \eg\ events obtained using the entire data set with
cuts applied in the order described above.
The irreducible background consists almost entirely of 
\epair\  events with one radiated photon.

\subsection{Kinematic Fits}
\label{section:kinfit}
The existence of excited leptons would reveal itself as an excess in
the total number of observed events, appearing as a peak in the 
reconstructed $\ell\gamma$ invariant mass distributions.
Kinematic fits are used to improve the reconstructed mass resolution
of the selected events and also further reduce the background thereby
increasing the sensitivity of the analysis to excited leptons. 

The kinematic variables used as input to the fit are the
energy and direction ({\it E}, $\theta$, $\phi$) of each
identified jet in an 
event.  The energy of tau candidates is left as a free parameter and
the direction is taken to be the jet axis.
A kinematic fit also requires as input
the error on each measured variable.
Estimates of the uncertainties on the energy and direction for
the different types of leptons and for photons are obtained from
studies of di-lepton events in data recorded at \cm\ energies near
and greater than the \z\ mass.
The error estimates are parameterised as functions of the jet
energy and polar angle.  The uncertainty on the jet energy is typically
2~GeV for electrons and photons, and about 5~GeV for muon candidates.
The uncertainty on the jet polar angle is about 2~mrad for electrons
and muons, 4~mrad for photons, and 7~mrad for tau
candidates.  Finally, the azimuthal angle of electron,
muon, photon and tau candidates is typically known to 0.4~mrad,
0.4~mrad, 3.5~mrad and 7~mrad respectively.

The kinematic fit enforces conservation of energy and momentum while
taking into account the beam energy spread as measured by the LEP
energy working group~\cite{bib:LEP_energy}.  This last constraint is
necessary since the expected mass resolution for excited leptons is of the same
order as the \cm\ energy spread, which is measured to be approximately 250~MeV.

Slightly different kinematic fits are applied for single and pair
production candidate events.
In addition to energy and momentum conservation, the 
kinematic fits for pair produced excited lepton candidates
also require the invariant masses of the two lepton-photon pairs in
the event to be
equal.  There are two possible lepton-photon
pairings in each event, and for each pairing an additional fit is performed
assuming the presence of an undetected initial state radiation 
photon along the beam axis.
Thus for each pair production candidate, 
four kinematic fits are performed.  
Similarly, two kinematic fits are applied to singly produced excited
lepton candidate events.   In the first case, only the  
two leptons and one photon are included in the fit.  
In the second case, the fit is performed assuming the presence of an 
initial state radiation photon along the beam axis.
Finally, a single kinematic fit is performed for $\rm e\gamma$
events.  The fit assumes the presence of an undetected electron
along the beam axis.

For a given final state, events are rejected if every kinematic fit
attempted has a probability less than 0.001.
When more than one successful kinematic fit is obtained for an event,
the fit performed without the presence of an initial state radiation
photon is chosen if the fit probability is greater than 0.001. 
For pair production candidates where two fits without initial state
radiation are performed, the lepton-photon pairing
corresponding to the fit with the highest probability in excess of
0.001 is chosen.  Otherwise, results from the fit with an
initial state radiation photon are retained where, for pair
production candidates, the lepton-photon pairing
corresponding to the fit with the highest probability is chosen.
This requirement on the kinematic fit probabilities reduces the
number of selected events by more than 70\% for \llgg\ final states
and by about 10\% for final states compatible with the single
production of an excited lepton.

\begin{figure}
\begin{center}
\mbox{\epsfig{file=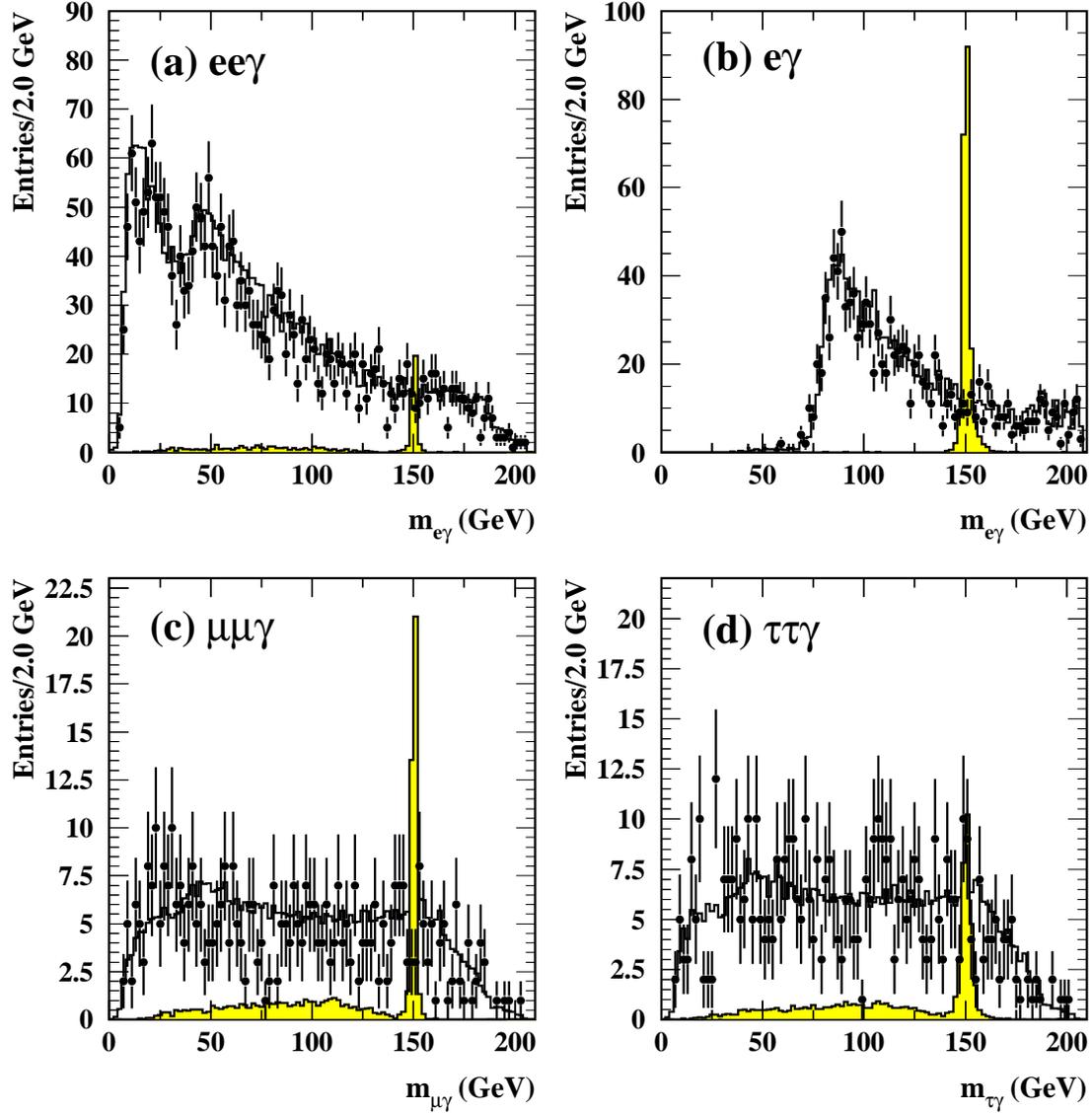,height=17cm}}
\caption{\label{fig:mlg} Reconstructed invariant mass distributions
for (a) \eeg, (b) \eg, (c) \mmg\ and (d) \ttg\ candidates after all
cuts are applied.  The points are data 
and the solid lines represent the total expected background from
Standard Model processes. 
The shaded histograms represent excited lepton signal events with an
arbitrarily chosen mass of 150~GeV and normalised to a ratio of the
coupling constant to the
compositeness scale of 0.4~$\rm TeV^{-1}$ (a,b) and 2~$\rm TeV^{-1}$ (c,d). 
There are two entries per event in (a,c,d) corresponding to the two possible
$\ell\gamma$ pairings.} 
\end{center}
\end{figure}

Results from the chosen kinematic fit for each event are used to calculate 
$\ell\gamma$ invariant masses.  
Using the procedure outlined above, the correct lepton-photon pairing for pair
produced excited leptons is chosen more than 98\% of the time as
determined using simulated signal events.
For \llg\ events, two lepton-photon combinations are possible, both of
which are included in the analysis.
Mass resolutions of approximately 0.2-0.4~GeV for
excited electrons and muons and 0.7-2.0~GeV for excited taus are
obtained using results from the kinematic fits.
The natural decay width of excited leptons for couplings not excluded
by previous searches is constrained to be much smaller than
these mass resolutions and is thus neglected.
Figure~\ref{fig:mlg} shows the invariant mass
distributions of selected \llg\ and \eg\ events.
There are two entries per \llg\ event,
corresponding to the two possible lepton-photon pairings.
These distributions are obtained by combining data from all the \cm\
energies considered. 
In both single and pair production event samples, no mass peaks are
observed in the data.

\section{Results}

The numbers of events observed in the data and the corresponding numbers of 
background events expected from Standard Model processes are shown in
Table~\ref{table:results}.
Selected candidate events for the pair production of excited leptons
are listed in Table~\ref{table:pair_candidates}.
Typical selection efficiencies for the pair production of excited
leptons vary from about 35\% to 55\%.
The efficiency for the single
production of excited muons is 70\% and approximately constant over
the entire kinematically allowed range of masses.
Near the kinematic limit for the single production of excited taus,
the efficiency rapidly drops from 53\% down to
approximately 20\% since the 
recoiling tau has low energy and thus often fails the initial set of
selection criteria.
For singly produced excited electrons, the efficiencies of the \eeg\ and \eg\
selections depend strongly on the mixture of $s$-channel and $t$-channel
components.   The sum of the \eeg\ and \eg\ efficiencies is typically
between 50\% and 70\%.  
The evaluation of the systematic uncertainties on the selection
efficiencies and background estimates is discussed
in the following section.  No excess of data indicating the existence
of excited leptons is found in either the single or pair production search.

%
%
\begin{table}
\begin{center}
\begin{tabular}{|l|r|r|r|r|r|r|r|} \cline{2-8}
\multicolumn{1}{c|}{} & \multicolumn{1}{c|}{$\rm ee\gamma$} & 
	\multicolumn{1}{c|}{$\rm e\gamma$} &  $\mu\mu\gamma$ &  $\rm
       \tau\tau\gamma$ & $\rm ee\gamma\gamma$ &  $\mu\mu\gamma\gamma$
       &  $\tau\tau\gamma\gamma$ \\ \hline\hline
Data & 1172 & 1123 & 212 & 248 & 3 & 3 & 7 \\ \hline
Background  & 1283  & 1229 & 239  & 260  & 4.0
        & 2.0  & 8.0 \\ \hline
Background statistical errors &  11 &  14 &  2 &  2 &  0.6 &
        0.2 &  0.6 \\ \hline
Background systematic errors & 161  & 159  & 28 & 40  & 1.2 & 0.7  & 6.3 \\ \hline
\multicolumn{8}{c}{} \\ \hline 
\multicolumn{8}{|c|}{Sources of background systematic errors} \\ \hline
ISR modelling & 141 & 135 & 26 & 29 & 0.3 & 0.1 & 0.6 \\ \hline 
Error estimate of fit variables & 68 & 57 & 10 & 24 & 1.1 &
       0.6 & 6.2 \\ \hline
Jet classification & 36 & 27 & 3 & 12 & 0.2 & 0.1  & 0.5 \\ \hline
Energy and angular resolution & 9 & 16 & 2 & 2 & 0.0 & 0.3
& 0.6 \\ \hline
Modelling of selection variables & 6 & 54 & 2 & 8 & 0.3 & 0.1 & 0.3 \\ \hline
\end{tabular}
\end{center}
\caption{\label{table:results} 
Total numbers of selected events in the data and 
expected numbers of background events for the different final
states considered.  
Statistical and total systematic uncertainties on the background
estimates are also shown.  Contributions to the total systematic
errors on the background expectations are listed in the lower part of
the table.}
\end{table}

%
%
\begin{table}
\begin{center}
\begin{tabular}{|c|c||c|c||c|c|} \hline
 \multicolumn{2}{|c||}{\eegg}  & 
                       \multicolumn{2}{c||}{\mmgg} & 
                       \multicolumn{2}{c|}{\ttgg}  \\ \hline
 \rule{0.cm}{0.4cm}\roots & $m_{\rm e\gamma}$ & 
                       \roots & $m_{\rm \mu\gamma}$ & 
                       \roots & $m_{\rm \tau\gamma}$  \\ 
 (GeV) & (GeV) & (GeV) & (GeV) & (GeV) & (GeV)
                       \\ \hline\hline
 188.7 & 39.3 & 188.7 & 44.5 & 188.6 & 70.9 \\ \hline
 199.6 & 80.1 & 205.1 & 28.0 & 189.0 & 52.7 \\ \hline
 201.6 & 92.7 & 206.2 & 34.0 & 199.6 & 30.1 \\ \hline
 \multicolumn{4}{c|}{ }      & 199.7 & 76.2 \\ \cline{5-6}
 \multicolumn{4}{c|}{ }      & 204.7 & 39.1 \\ \cline{5-6}
 \multicolumn{4}{c|}{ }      & 204.8 & 38.6 \\ \cline{5-6}
 \multicolumn{4}{c|}{ }      & 205.1 & 88.6 \\ \cline{5-6}
\end{tabular}
\end{center}
\caption{\label{table:pair_candidates} List of selected candidate
events for the pair production of excited leptons.  For each
candidate, the \cm\ energy and reconstructed invariant mass obtained
after performing the kinematic fits are listed. }
\end{table}

\subsection{Systematic Uncertainties}
The following sources of systematic uncertainties on the signal
efficiencies and background estimates were investigated.  These are
described in order of importance.

%
%
Uncertainties in the modelling of initial state photon radiation (ISR) in
di-lepton events affect the background estimates.  They are
assessed by comparing background expectations from 
the KORALZ and KK2F event generators for the processes $\rm
e^+e^-\rightarrow\mu^+\mu^-$ and $\rm
e^+e^-\rightarrow\tau^+\tau^-$.  The Monte Carlo program KK2F, used in
this analysis to estimate the background contributions from
$\mu^+\mu^-$ and $\tau^+\tau^-$ events, has the most
complete description of initial state radiation including second-order
subleading corrections and the exact matrix elements for two hard
photons~\cite{bib:ceex}.  
The relative variations in background expectations between the two
Monte Carlo generators are assigned as
systematic uncertainties representing the effect of missing higher
orders.  These are found to be 11\% for final states compatible with
the single production of excited muons and taus, and 7\% for \mmgg\ and
\ttgg\ events. 
The BHWIDE and TEEGG event generators, used to simulate
the background 
from radiative \epair\ events, have a precision for radiative
corrections similar to the KORALZ program.  The
background estimates for events expected from the
production of excited electrons are thus assigned an uncertainty of
7\% for the \eegg\ final state and 11\% for both \eeg\ and \eg\
events.  This uncertainty is
significantly larger than the error for inclusive electron pair
production cited in \cite{bib:lepmc}.

%
%
For the purpose of calculating limits on the product of the cross-section
and the branching fraction, it is necessary to be able to calculate the
efficiency and mass resolution of signal events at arbitrary excited
lepton masses and \cm\ energies.  
For each final state, the selection
efficiencies and mass resolutions are parameterised as a function of
the excited lepton mass scaled by the \cm\ energy ($m_{\ell^*}/\sqrt{s}$).
The systematic uncertainties associated with the interpolation of
efficiencies and mass resolutions were estimated by
calculating the root-mean-square spread between
simulated signal event samples and the parameterisation functions.

%
%
Uncertainties on the fit variable error estimates are evaluated by
varying the errors on each variable independently.  
The errors are varied by an amount
representing one standard deviation as calculated from the
uncertainties on the energy and angular parameterisation.
Background estimates for final states containing two leptons and two
photons are particularly sensitive to changes in the errors due
to the additional constraint in the kinematic fit requiring events to
have equal 
reconstructed lepton-photon invariant masses.  
Also, the smaller sample of tau
pair events used to parameterise the errors on the tau direction 
results in larger statistical uncertainties on the error
parameterisation which in turn 
dictate the larger variations used to estimate the systematic error
contributions.

%
%
The jet classification into leptons or photons 
contributes to the overall
systematic uncertainty through the modelling of the lepton and photon 
identification efficiencies.  
Using di-lepton and di-photon events recorded at \cm\ energies equal
to and greater than the \z\ mass, the systematic uncertainty
associated with each set of 
lepton and photon requirements was evaluated by comparing the
identification efficiencies obtained from data and simulated events.
Relative errors of 1\% for electron and muon, and 2\% for the tau and
photon classifications are assigned.
Systematic uncertainties associated with each final state were
determined by adding linearly contributions from identical jet 
classifications
and adding in quadrature contributions from different types of leptons
and photons.  The resulting uncertainties on the signal efficiencies,
shown in Table~\ref{table:syst_eff}, are fully correlated with the
corresponding errors on the background estimates presented in
Table~\ref{table:results}.

%
%
The systematic uncertainty associated with the energy scale, energy resolution
and angular resolution of the leptons and photons was evaluated by
modifying each parameter independently in Monte Carlo simulated
events.  
Comparisons between data and simulated distributions of di-lepton
events recorded at different \cm\ energies were used to determine the
size of these variations.
The energy (momentum) of electron and photon (muon) candidates was
shifted by 0.3\%.  The energy and angular resolutions of jets were
smeared by the maximum values for which simulated events were
compatible with the distribution of data within one standard deviation.
Variations in the energy scale result in negligible changes in
efficiencies and background.  Contributions to the systematic
uncertainty of each final state from individual changes in the energy
and angular 
resolution are added in quadrature.

%
%
The systematic uncertainty due to Monte Carlo modelling of the event selection
variables was estimated by varying each selection cut independently and measuring
the corresponding changes in the overall signal efficiencies and
background estimates.
The difference between the mean value of the data and background
expectation for each selection variable determined the range of
variation of each cut. 
Systematic uncertainties varying between 0.5\% and 6.3\% are assigned
to the different background estimates.  Contributions to the systematic
error on the signal
efficiencies are shown in Table~\ref{table:syst_eff}.

Lastly, the uncertainty on the integrated luminosity measurements
(0.2\%) is considerably smaller than the systematic effects already described
and is therefore neglected.

Summaries of the systematic effects on the background expectations and
signal efficiencies are presented in
Tables~\ref{table:results} and \ref{table:syst_eff}, respectively.
These systematic uncertainties are included in the
calculation of limits as described in the following section.

%
%
\begin{table}
\begin{center}
\begin{tabular}{|l|r|r|r|r|r|r|r|}\hline
\multicolumn{1}{|c}{Source} & \multicolumn{7}{|c|}{Uncertainty (\%)} \\ \cline{2-8}
       &$\rm ee\gamma$ & $\rm e\gamma$ &  $\mu\mu\gamma$ &  $\rm
       \tau\tau\gamma$ & $\rm ee\gamma\gamma$ &  $\mu\mu\gamma\gamma$ &  
$\tau\tau\gamma\gamma$ \\ \hline\hline
Resolution interpolation & 18.6 & 12.5 & 20.7 & 7.1 & 23.5 & 18.1 &
       12.9 \\ \hline
Efficiency interpolation  & 8.6 & 3.0 & 2.3 & 5.3 & 4.5 &
3.4 & 4.4 \\ \hline
Error estimate of fit variables & 5.0 & 5.0 & 3.0 & 5.0 & 3.0 & 2.0 & 6.0
 \\ \hline
Jet classification & 2.8 & 2.2 & 1.4 & 4.5 & 4.5 & 4.5  & 5.7 \\ \hline
Energy and angular resolution & 0.9 & 1.1 & 0.8  & 1.1  & 1.8  & 0.6 & 0.9 \\ \hline
Modelling of selection variables & 0.0 & 1.5  & 0.1 & 1.6 & 0.4 & 0.4 & 0.8 \\ \hline\hline
Total & 21.3 & 14.1 &  21.1 & 11.3 & 24.6 & 19.1 & 16.0  \\ \hline
\end{tabular}
\end{center}
\caption{\label{table:syst_eff} Relative systematic uncertainties on the signal
efficiencies for each final state considered.  }
\end{table}

\subsection{Limit Calculations}
Limits on the product of the cross-section and the electromagnetic
branching fraction of excited leptons are obtained from both pair and
single production searches.
The numbers of data and expected
background events at each \cm\ energy are binned as a function of the 
reconstructed invariant mass.  For selected \llg\ candidates, both possible
$\ell\gamma$ invariant masses are used.  Due to the excellent mass resolution, the double-counting of \llg\ events does not affect the limits calculated.
Each mass bin at a given \cm\ energy is
treated as an independent counting experiment.  

For the purpose of calculating limits, the
signal invariant mass is assumed to be well described by a Gaussian 
distribution centred
at the test mass value and with a width equal to the expected mass 
resolution.  
The validity of this assumption is verified with Monte Carlo
simulation of signal events at different masses and \cm\ energies.
Efficiency corrections due to non-Gaussian tails in the
invariant mass distributions are applied to the signal expectation.
These correction factors, signal efficiencies and mass resolutions 
 are all parameterised as a function of the excited lepton mass scaled by
the \cm\ energy, $m_{\ell^*}/\sqrt{s}$. 
The efficiency correction factors are constant over the entire
kinematically allowed range and vary from
approximately 0.7 to 0.85 depending on the event final state.
Efficiencies and mass resolutions are well-described by polynomial
functions of various degrees.
The efficiencies for the single production of excited leptons are
calculated with the production and decay angular distributions corresponding to
$f=f^\prime$.  The assignment $f=f^\prime$ particularly affects the
relative fraction of excited 
electron events in the \eg\ and \eeg\ selections.

For a given test mass,  the Gaussian distributions
describing the invariant mass of signal events at each \cm\ energy 
considered are normalised 
to the expected excited lepton cross-section at the highest \cm\
energy, thereby taking into account the energy dependence of the cross-section.
A likelihood ratio
method~\cite{bib:likelihood_ratio} is used to compute the 95\%
confidence level upper limit on the 
number of signal events produced in the entire data set ($\rm N_{95}$).
Systematic uncertainties on the signal efficiency and background
expectation are incorporated 
by fluctuating, over many iterations, the background expectation and
signal efficiency according to their respective systematic
uncertainties.  The final limits are determined from the
average of all the $\rm N_{95}$ values obtained at each iteration.
Systematic errors on the background estimates
are treated as being fully correlated.  
The systematic uncertainties on the signal efficiencies due 
to the jet classification are also fully correlated with the corresponding 
errors on the background estimates and are treated as such in the limit 
calculations.

\begin{figure}
\begin{center}
\mbox{\epsfig{file=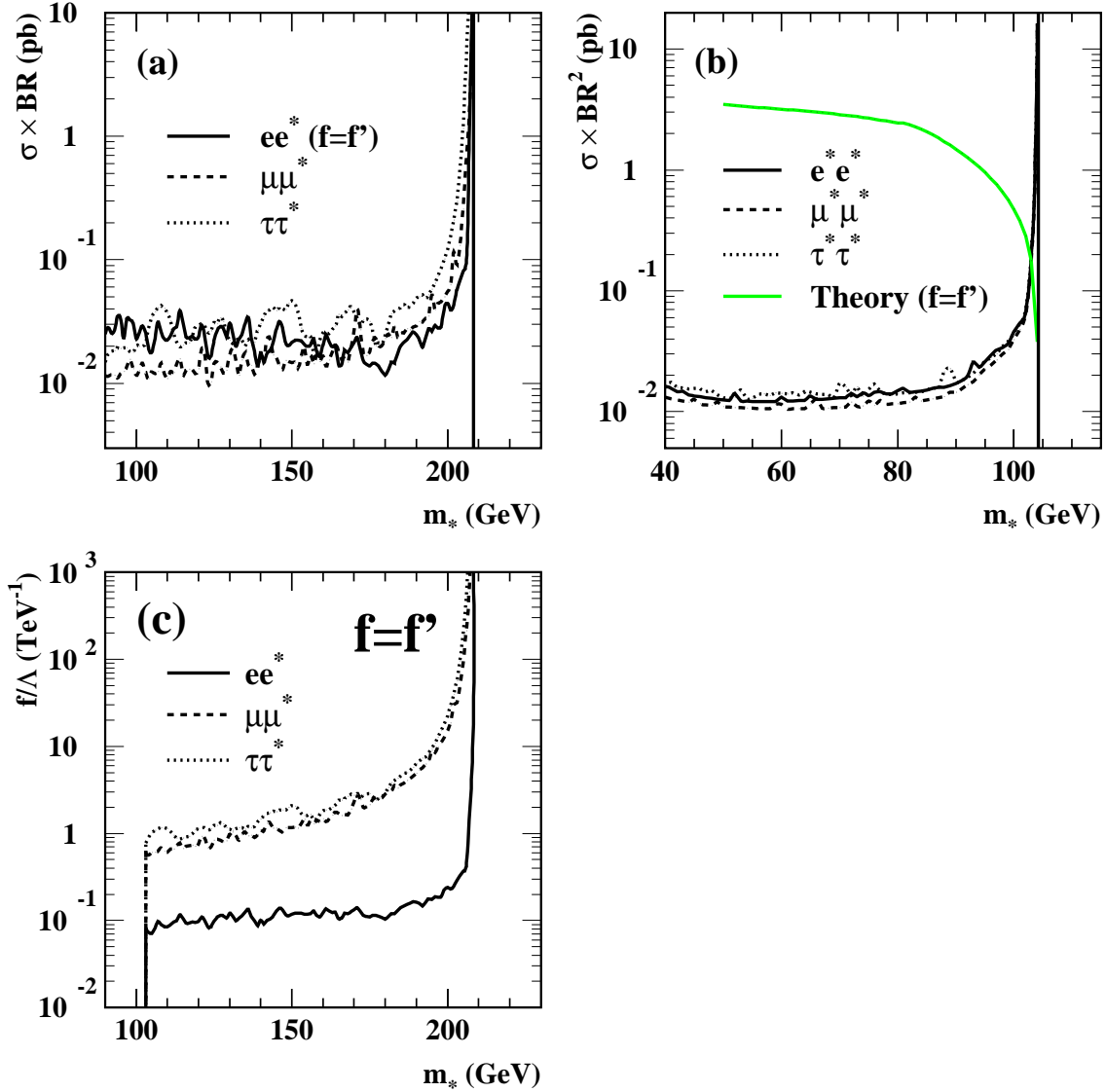,height=17cm}}
\caption{\label{fig:limits} The 95\% confidence level upper limits 
on the product of the cross-section at \sqrts~=~\lastcm~GeV and the branching
fraction for (a) single and (b) pair production of excited leptons as
a function of mass ($m_*$).  The limit obtained for the single
production of excited electrons 
is calculated assuming $f = f^\prime$.  The regions above the curves
are excluded.  The product of the theoretical cross-section and the branching
fraction squared assuming $f=f^\prime$ is also shown in (b).
The 95\% confidence level upper limits 
on the ratio of the excited lepton coupling constant to the
compositeness scale, $f/\Lambda$, as 
a function of the excited lepton mass and assuming $f=f^\prime$ are
shown in (c).  The regions above the curves are 
excluded by single production searches while
pair production searches exclude masses below \meegg~GeV for excited
electrons, muons and taus.}
\end{center}
\end{figure}

Limits on the product of the cross-section and the
branching fraction are scaled to \sqrts~= \lastcm~GeV assuming the
cross-section evolution as a function of \cm\ energy expected for
excited leptons.  
The upper limits on the single production of
excited muons and tau leptons do not depend on the coupling assignment
of $f$ and $f^\prime$.
The excited electron selection efficiencies, however, 
depend on the relative magnitude of
the {\it s}-channel and {\it t}-channel diagrams.  For comparison with
previously published results, the limits on excited electrons presented
here assume $f=f^\prime$.  Figures~\ref{fig:limits}(a,b) 
show the 95\% confidence level upper limits on the
product of the cross-section at \sqrts~= \lastcm~GeV and the branching
fraction obtained from the search for singly and pair produced excited leptons.

The theoretical calculation~\cite{bib:ex_theory1} of the product of
the pair production cross-section at 
\sqrts~= \lastcm~GeV and the branching fraction squared is
overlayed on Figure~\ref{fig:limits}(b). 
As part of this calculation, the electromagnetic branching fraction is 
calculated assuming $f=f^\prime$.
The 95\% confidence level lower mass limits on excited leptons correspond 
to the mass at which the cross-section times branching fraction limit curves
cross the theoretical expectation.  Lower mass limits of $m_{e^*}
>$~\meegg~GeV, $m_{\mu^*} >$~\mmmgg~GeV and $m_{\tau^*} >$~\mttgg~GeV are
obtained.  Although systematic
errors are incorporated into the limit calculations, an
additional uncertainty on the mass limits arises from the finite width
of the \cm\ energy bins considered.  The 0.5~GeV \cm\ energy bin width
near the kinematic limit corresponds to an uncertainty of  
0.1~GeV on the mass limits.

Limits on the product of the cross-section and the electromagnetic
branching fraction of singly 
produced excited leptons are used to constrain parameters of the model
introduced in Section~\ref{section:intro}.
Since the cross-section for the single production of excited leptons
is proportional to $(f/\Lambda)^2$, limits on the ratio of the coupling to
the compositeness scale as a function of excited lepton mass are
calculated using
\[
  \left[\frac{(f/\Lambda)}{(1\ \rm TeV^{-1})}\right]_{\rm 95\% CL} =
  \sqrt{\frac{\rm N_{95}}{\rm N_{\rm exp}}}\:\:\: ,
\]
\noindent
where $\rm N_{\rm exp}$ is the number of expected signal events
assuming $f/\Lambda = 1~\rm TeV^{-1}$.
Figure~\ref{fig:limits}(c) shows these limits for each type of excited
lepton.
The $f/\Lambda$ limit for excited electrons is
approximately an order of magnitude better than for muons and taus due to
the enhancement of the cross-section coming from the {\it t}-channel contribution.

\section{Conclusion}

A search for electromagnetically decaying charged excited leptons was
performed using 684.4~$\rm pb^{-1}$ of data collected by the OPAL detector at
\sqrts\ = 183-209 GeV.  No evidence was found for the existence of
excited leptons. 
Upper limits on the product of the cross-section and the branching
fraction were calculated.   
From pair production searches, 95\% confidence level lower limits on
the mass of excited leptons
are determined to be $m_{\ell^*}$~$>$~\meegg~GeV for $\ell = \rm
e,\mu, \tau$.
From the results of the search for singly produced
excited leptons, limits were calculated on the ratio of the coupling constant
to the compositeness scale ($f/\Lambda$) as a function of excited lepton mass.
The results are currently the most stringent constraints on the
existence of excited 
leptons and therefore represent a significant improvement on limits
previously published~\cite{bib:ex_opal_results,bib:lep,bib:hera}.

%
%
\bigskip\bigskip
\noindent
{\large\bf Acknowledgements} \\
\par
We particularly wish to thank the SL Division for the efficient operation
of the LEP accelerator at all energies
 and for their close cooperation with
our experimental group.  In addition to the support staff at our own
institutions we are pleased to acknowledge the  \\
Department of Energy, USA, \\
National Science Foundation, USA, \\
Particle Physics and Astronomy Research Council, UK, \\
Natural Sciences and Engineering Research Council, Canada, \\
Israel Science Foundation, administered by the Israel
Academy of Science and Humanities, \\
Benoziyo Center for High Energy Physics,\\
Japanese Ministry of Education, Culture, Sports, Science and
Technology (MEXT) and a grant under the MEXT International
Science Research Program,\\
Japanese Society for the Promotion of Science (JSPS),\\
German Israeli Bi-national Science Foundation (GIF), \\
Bundesministerium f\"ur Bildung und Forschung, Germany, \\
National Research Council of Canada, \\
Hungarian Foundation for Scientific Research, OTKA T-029328, 
and T-038240,\\
Fund for Scientific Research, Flanders, F.W.O.-Vlaanderen, Belgium.\\

\pagebreak

%
%

%
%

%
%

\end{document}